# Effects of W alloying on the electronic structure, phase stability and thermoelectric power factor in epitaxial CrN thin films

Niraj Kumar Singh,[1,2*] Victor Hjort,[1] Sanath Kumar Honnali,[1] Davide Gambino,[3] Arnaud le Febvrier,[1] Ganpati Ramanath,[1,4] Björn Alling[3] and Per Eklund[1,2]

[1]*Thin Film Physics Division, Department of Physics, Chemistry, and Biology (IFM), Linköping University, SE-581 83, Linköping, Sweden*

[2] *Inorganic Chemistry, Department of Chemistry - Ångström Laboratory, Uppsala University, Box 538, SE-751 21 Uppsala, Sweden*

[3]*Theoretical Physics Division, Department of Physics, Chemistry, and Biology (IFM), Linköping University, Linköping-58183, Sweden*

[4]*Rensselaer Polytechnic Institute, Materials Science & Engineering Department, Troy, NY 12180, USA*

*E-mail: niraj-kumar.singh@kemi.uu.se

**Abstract**

CrN-based alloy thin films are of interest as thermoelectric materials for energy harvesting. *Ab initio* calculations show that dilute alloying of CrN with 3 at.% W substituting Cr, induce flat electronic bands and push the Fermi level $E_F$ into the conduction band, while retaining dispersive Cr 3*d* bands. These features are conducive for both high electrical conductivity $\sigma$ and high Seebeck coefficient $\alpha$, and hence the thermoelectric power factor $\alpha^2\sigma$. To investigate this possibility, epitaxial $CrW_xN_z$ films were grown on *c*-plane sapphire by dc-magnetron sputtering. However, even films with the lowest W concentration (x = 0.03) in our study contained metallic *h*-$Cr_2N$, which is not conducive for a high $\alpha$. Nevertheless, the films exhibit a sizeable power factor of $\alpha^2\sigma \sim 4.7 \times 10^{-4}$ $Wm^{-1}K^{-2}$ due to high $\sigma \sim 700$ $Scm^{-1}$, and a moderate $\alpha \sim -25$ µV/K. Increasing *h*-$Cr_2N$ fractions in the $0.03 < x \leq 0.19$ range monotonically increases $\sigma$, but severely diminishes $\alpha$ leading to two orders of magnitude decrease in $\alpha^2\sigma$. This trend continues with x > 0.19 due to W precipitation. These findings indicate that dilute W additions below its solubility limit in CrN is important for realizing high thermoelectric power factor in $CrW_xN_z$ alloy films.

## I. Introduction

Transition metal nitrides (TMNs) constitute an important class of materials that exhibit a combination of high hardness, mechanical strength, thermal stability and corrosion resistance [1]. TMNs such as TiN, HfN and ZrN have been used as protective and decorative coatings, and nanolayers of metallic TMNs are used as diffusion barriers in nanodevice wiring architectures in integrated circuits [2–5]. Narrow-bandgap thin films of TMNs, such as, CrN- and ScN-based alloys, have been considered for thermoelectric and piezoelectric energy conversion [6–9]. ScAlN films exhibit large piezoelectric responses at 500 °C making them suitable for high-temperature device applications [10–12].

Thermoelectric conversion efficiency is related to a dimensionless figure of merit $ZT = \frac{\alpha^2 \sigma T}{\kappa}$, where T is the absolute temperature, $\alpha$ is the Seebeck coefficient, $\sigma$ the electrical conductivity, $\kappa$ the thermal conductivity [13,14] and $\alpha^2\sigma$ is the thermoelectric power factor. Realizing high *ZT* is a fundamental challenge that requires the circumvention of unfavorable couplings between $\alpha$, $\sigma$ and $\kappa$. *ZT* enhancement strategies include doping (less than a few at.%) [15] and/or alloying (several at.% or more) [16] to increase $\alpha^2\sigma$ by manipulating the electronic structure, and decreasing $\kappa$ by impurity- and nanostructuring-induced phonon scattering [17,18].

CrN-based alloys have inherently high thermoelectric power factors [19–21] due to the narrow bandgap of CrN ($E_g$ ~ 0.2 eV) which simultaneously induces high $\sigma$ and high $\alpha$ underpinned by high charge carrier entropy near the Fermi level $E_F$. These attributes, together with an intrinsically low $\kappa$ ~ 2 $Wm^{-1}K^{-1}$ when compared to other TMNs [22], due to low phonon lifetimes arising from spin-lattice coupling [23,24], make CrN an attractive candidate for enhancing *ZT* through doping and alloying. Substituting Cr ($3d^5$) with W ($5d^4$) in CrN offers the possibility to not only manipulate σ and α, but also decrease κ due to the atomic mass contrast between W and Cr. Indeed, adding W or V to CrN increase $α^2σ$, but *ZT* increases only for W additions which decrease κ, [25] unlike V additions which increase κ [26]. Understanding the effects of W alloying on the electronic structure, phase stability and film morphology are crucial for understanding and tuning the thermoelectric properties to realize high *ZT* in CrN-based films.

Here, we show from *ab initio* calculations that dilute substitutions of Cr with W in CrN result in flat electronic bands near the Fermi level $E_F$, which is pushed into the conduction band. These features are conducive for high σ as well as high α, and hence, a high power-factor $α^2σ$. To examine this experimentally, we attempted to grow epitaxial $CrW_xN_z$ films, and measured σ and α as a function of W content x. We find that W additions monotonically increase σ, but severely diminish α due to increasing fractions of metallic *h*-$Cr_2N$ and W precipitation. These findings indicate that restricting W content to below its solubility limit in CrN is crucial for realizing high $α^2σ$ in $CrW_xN_z$ alloys.

## II. Methods

### A. *Computational details*

The electronic structures of cubic $Cr_{1-x}W_xN$ with x = 0 and x = 0.03 in double layer antiferromagnetic (AFM $[110]_2$) configuration were computed by density functional theory calculations (DFT) using projector-augmented-wave (PAW) potentials as implemented in the Vienna *ab initio* simulation package (VASP) code [27,28]. The exchange-correlation energy was treated within the LDA + U scheme with *U* = 3 eV [29] to consider Coulombic interactions in the narrow Cr 3*d* bands, and counteract bandgap underestimation or absence by local approaches. The cutoff energy of plane waves for the expansion of wavefunctions was set to

520 eV. Brillouin zone sampling was carried out on a 5×5×5 Monkhorst-Pack $k$ mesh, and the electronic optimization criterion was set to $10^{-5}$ eV. A 2×2×2 supercell of 64 atoms was used to analyze the equation of state (EOS) to estimate the equilibrium lattice constant $a_0$. The crystal structure was visualized using Vesta [30] and the electronic structures were plotted using Sumo package [31].

## B. *Experimental details*

CrW$_x$N$_z$ thin films with $0 \leq x \leq 0.25$ were grown in an ultra-high vacuum chamber with a base pressure of ~$10^{-6}$ Pa [29] by DC magnetron sputtering onto 10 x 10 mm$^2$ $c$-plane sapphire substrates (Alineason Materials Technologies). Prior to film growth, the substrates were ultrasonicated successively in acetone and ethanol for 10 minutes each, followed by blow drying with a N$_2$ gas jet. The films were sputter-deposited from 50-mm-diameter disc-shaped targets of Cr and W (both 99.95% purity) using a 0.33 Pa (~ 2.5 mTorr) plasma struck with a 40/60 Ar/N$_2$ mixture. The W level was varied by fixing the Cr target power $P_{Cr}$ constant and varying the W target power $P_W$ in the $0.1 \leq P_W/P_{Cr} \leq 0.65$ range. During film deposition, the substrates were at 600 °C and rotated at 15 rpm.

Phase purity and crystal structure were characterized by X-ray diffractometry (XRD) using a PANanlytical X'Pert Pro instrument with a Cu-K$_\alpha$ radiation source operated at 45 kV and 40 mA. Pole figure measurements were carried out by scanning along in-plane rotation ϕ and out-of-plane tilt ψ angles at 2.5° steps per second using cross-slit primary optics and parallel plate collimator secondary optics. Film thickness, roughness, and density were estimated by analyses of X-ray reflectivity (XRR) data acquired using a hybrid-mirror module with a 0.5° incidence slit and a 0.125° post-diffraction slit. Film morphology was examined by scanning electron microscopy (SEM) in a Sigma 300 Zeiss instrument operated at 3 kV using Inlens detector. Measurement of composition of the films was carried out by combination of Rutherford backscattering spectrometry (RBS) and time-of-flight elastic recoil detection analysis (ToF-ERDA). Experiments were performed using 5 MV 15-SDH2 Pelletron accelerator at Tandem Laboratory, Uppsala University [32]. 2 MeV He$^+$ primary ion beam was incident at 5° to the sample normal and the backscattered particles were detected at an angle of 170°. The acquired spectra was processed and analysed using the *SIMNRA* [33] software, version *7.03*. For ToF-ERDA measurements, 44 MeV iodine ($^{136}$I$^{8+}$) beam was incident at an angle of 67.5° with the sample surface normal. The incident ions created recoils which passed through the time-of-flight detector and gas ionization chamber, placed at an angle of 45° relative to the incident beam. The elemental depth profile was extracted from the time and energy coincidence spectra using *Potku* 2.0 code [34].

Room-temperature electrical conductivity σ was measured by using a Jandel Model RM3000 four-point-probe station. Seebeck coefficient, α was measured using a home-made setup equipped with Peltier devices that maintain desired temperature gradient across the sample. In this setup, temperatures are measured with K-type thermocouples, and the Seebeck voltage is measured with a Keithley 2001 multimeter as described elsewhere [35]. Charge carrier density n and mobility μ were measured by room-temperature Hall measurements in a home-made setup with a 0.485 T permanent magnet. The Hall voltage was measured in the van der Pauw configuration with a Keithley 2100 multimeter, with a 40-mA current supplied by a Keithley 2400 source.

## III. Results and discussion

### A. *Ab-initio calculations*

CrN undergoes a magnetostructural transition at the Néel temperature $T_\mathrm{N}$ ~ 280 K, wherein the antiferromagnetic low-temperature orthorhombic phase is converted to the high-temperature paramagnetic rocksalt phase [36]. However, cubic CrN is known to energetically favor the antiferromagnetic spin configuration with alternating double layers of identical spin perpendicular to the crystallographic [110] direction (AFM-$110_2$) as shown in Fig. 1a [33]. The equilibrium lattice parameter $a_0$ of cubic CrN was estimated by fitting energy (E) vs volume (V) to obtain $a_0$ ~ 4.13 Å for CrN (Fig. 1b), which agrees well with prior theoretical and experimental reports [38,39]. For $Cr_{1-x}W_xN$ with x = 0.03, we obtain $a_0$ ~ 4.14 Å for W occupancy of either Cr↑ or Cr↓ sites, suggesting no observable site-occupancy effect. Thus, we calculated the electronic properties only for W substitution at Cr↑ sites.

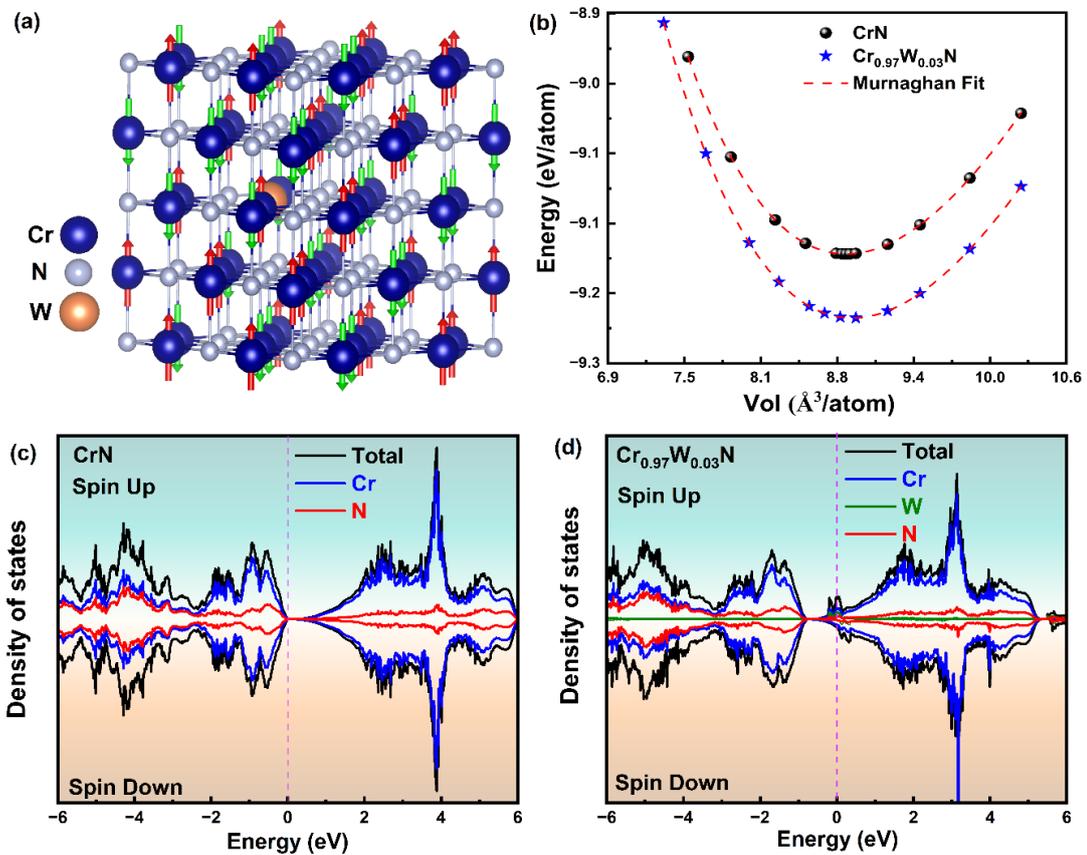

Fig. 1. (a) A sketch of the cubic CrN crystal structure showing Cr↑ (red) and Cr↓ (green) sites, with W substitution at the Cr↑ sites. (b) Murnaghan fits for CrN and $Cr_{0.97}W_{0.03}N$. Density of states plots for (c) CrN and (d) $Cr_{0.97}W_{0.03}N$.

For pristine defect-free CrN, the calculated bandgap $E_g$ ~ 0.2 eV agrees with prior work [37], and the Cr 3$d$ states are the primary contributors to the states near $E_F$ (see Fig. 1c).

For $Cr_{0.97}W_{0.03}N$, the $E_F$ is inside the conduction band (Fig. 1d) and is associated with a significantly higher density of states near $E_F$ than in CrN. This W-induced $E_F$ shift into the conduction band is likely to produce metal-like electrical transport, commonly observed in degenerate semiconductors.

Electronic structure calculations for CrN along the high symmetry X-Γ-K-W-X path reveal dispersive Cr 3d bands with conduction band minima at Γ, and a bandgap of $E_g$ ~ 0.2 eV (Fig. 2a). Substituting Cr with 3 at.% W, i.e., x = 0.03, largely preserves the CrN band structure, but results in additional flat bands with a strong W-d character near $E_F$. The simultaneous presence of dispersive and flat bands, and high density of states near $E_F$ are conducive for high $α^2σ$ in $Cr_{1-x}W_xN$ with x = 0.03. Dispersive bands favor high carrier mobility μ, and hence, high σ. Flat narrow bands favor high α due to the high effective mass m* as seen from $α = \frac{8π^2k_B^2}{3eh^2}m^*T\left(\frac{π}{3n}\right)^{2/3}$, where n is the charge carrier density [13]. Thus, higher m* tends to decrease σ leading to the inverse coupling between α and σ. Thus, the presence of both narrow as well as highly dispersive bands are necessary for realizing high $α^2σ$ [36].

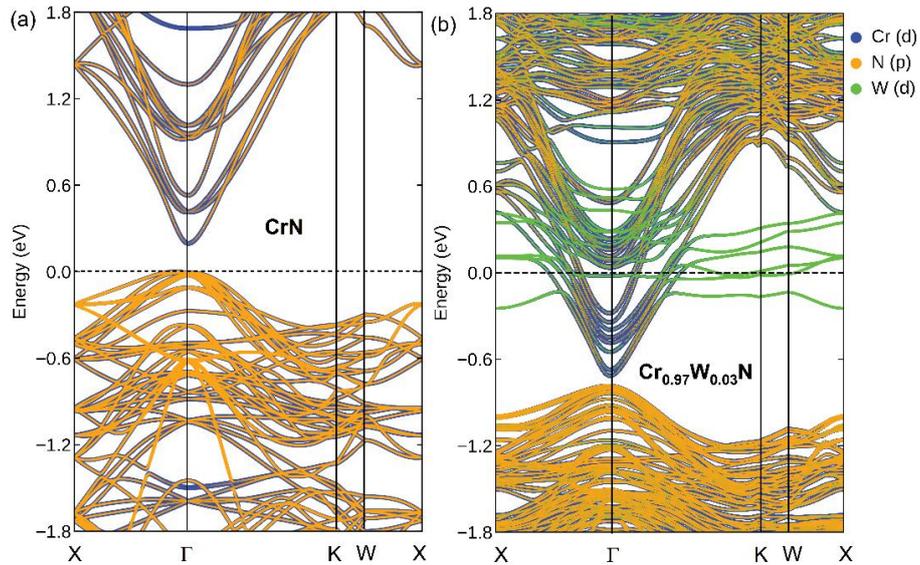

Fig. 2. Electronic band structures computed by *ab initio* calculations along X-Γ-K-W-X for (a) CrN and (b) $Cr_{0.97}W_{0.03}N$.

## B. Epitaxial CrW$_x$N growth and structural analysis

To experimentally investigate our theoretical predictions, we attempted to grow epitaxial Cr$_{1-x}$W$_x$N$_\delta$ films. We chose a growth temperature of 600 °C because epitaxial rocksalt CrN is favored in the 550 °C ≤ T ≤ 700 °C range [41], and metallic $h$-Cr$_2$N formation at higher temperatures is not conducive for high α. Cr target power P$_{Cr}$ is also a crucial parameter that influences CrN and $h$-Cr$_2$N phase selection and fractions, and Cr vacancy concentration that can impact σ [42]. Results of our studies of CrN films grown with 42 W ≤ P$_{Cr}$ ≤ 50 W showed that the highest α$^2$σ was obtained for P$_{Cr}$ = 44 (see supplementary information Fig. S1). Accordingly, CrW$_x$N$_z$ films were deposited for different P$_W$ at a fixed P$_{Cr}$ = 44 W.

RBS and ERDA analyses showed that all the CrW$_x$N$_z$ films were nitrogen-deficient (see Table I). Increasing P$_W$ in the 0.1 ≤ P$_W$/P$_{Cr}$ ≤ 0.65 range increased the W content from x = 0.03 to 0.25, decreased the nitrogen content from z = 0.91 to 0.21, and increased oxygen contamination from 1.7 at. % to 5.5 at.%. The large decreases in nitrogen content correlating with increasing W content suggests the formation of a Cr-rich Cr$_2$N, corroborated by XRD analyses described below.

Table I. Elemental compositions of CrWN films determined by RBS and ToF-ERDA.

| *Cr* | *W* | *N* | *O* | *N/(Cr+W)* |
|---|---|---|---|---|
| *51.6* | - | 46.9 | 1.7 | 0.91 |
| *50.5* | 3.4 | 43.4 | 2.8 | 0.80 |
| *58.4* | 6.8 | 30.6 | 4.2 | 0.47 |
| *55.8* | 18.9 | 21.0 | 4.3 | 0.28 |
| *53.0* | 25.1 | 16.4 | 5.5 | 0.21 |

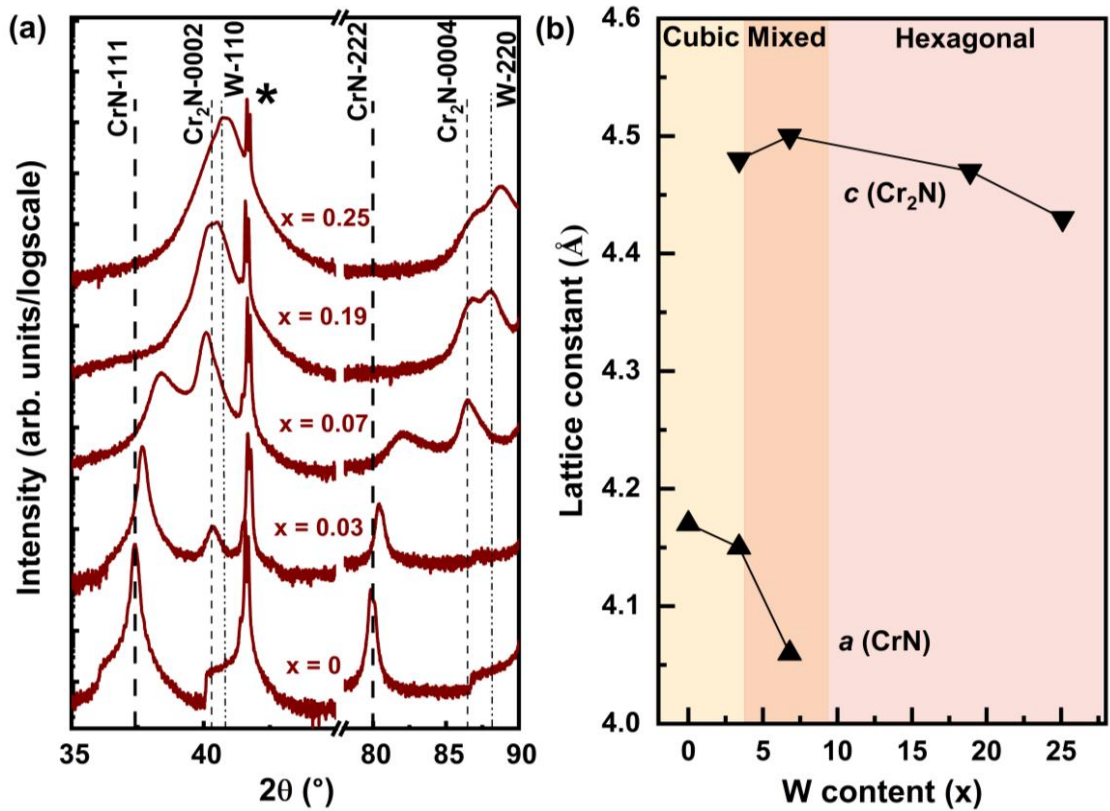

Fig. 3. X-ray diffractograms from (a) $CrW_xN_z$ films with $0.00 \leq x \leq 0.25$. Here, * represents the substrate 006 peak at $2\theta \sim 41.67°$ (b) Lattice parameters of CrN and $h$-$Cr_2N$ for $CrW_xN_z$ films with $0.00 \leq x \leq 0.25$.

X-ray diffractograms show that W additions shift the 111 and 222 CrN Bragg peaks to higher angles, i.e., lower d-spacings, indicative of W-induced unit cell compression. Even for $x \sim 0.03$, the lowest W content explored in this study, we observe the $h$-$Cr_2N$ (0002) reflection at $2\theta \sim 40.3°$. The results indicate that the solubility limit of W in CrN is less than $x \sim 0.03$ even under the far-from equilibrium conditions involved during sputter deposition. Our attempts to obtain CrN films with lower W contents were unsuccessful due to the instability of the plasma at low $P_W$. Thus, alternative approaches need to be implemented for achieving desired dilute W concentrations in CrN films.

At $x = 0.07$, the CrN peaks continue to shift further but considerably decrease in intensity, while the $h$-$Cr_2N$ 0002 peak intensity increases and the $h$-$Cr_2N$ 0004 peak appears at $2\theta \sim 86.5°$.

No rocksalt CrN peaks were detectable in diffractograms from films with x = 0.19, whereas the *h*-Cr$_2$N peaks were stronger. Films with x = 0.25 exhibit 110 and 220 Bragg peaks at 2θ ~ 40.55° and ~ 88.1° from metallic W, showing the coexistence of metallic *h*-Cr$_2$N and W.

X-ray pole figures of the 111 and 002 reflections from the CrN$_z$ and CrW$_{0.25}$N$_z$ films, respectively, show distinct spot patterns at specific angles (Fig. 4) confirming the epitaxial growth of CrW$_x$N$_z$ films. Scans of the {111} poles with Bragg angle 2θ = 37.4° show a central peak and 6 peaks at ψ ~ 71° (Fig. 4a). Here, the presence of 6 peaks rather than 3 expected is indicative of twin domains seen in epitaxial films of rocksalt structure materials grown on *c*-plane sapphire. Pole figure performed at 2θ = 73.1° on the CrW$_{0.25}$N$_z$ film reveals the presence of 6 poles at ψ ~ 65° with a symmetry of 3 (Fig. 4(b)). These poles correspond to the {03$\bar{3}$1} reflection from *h*-Cr$_2$N and demonstrating its epitaxial character.

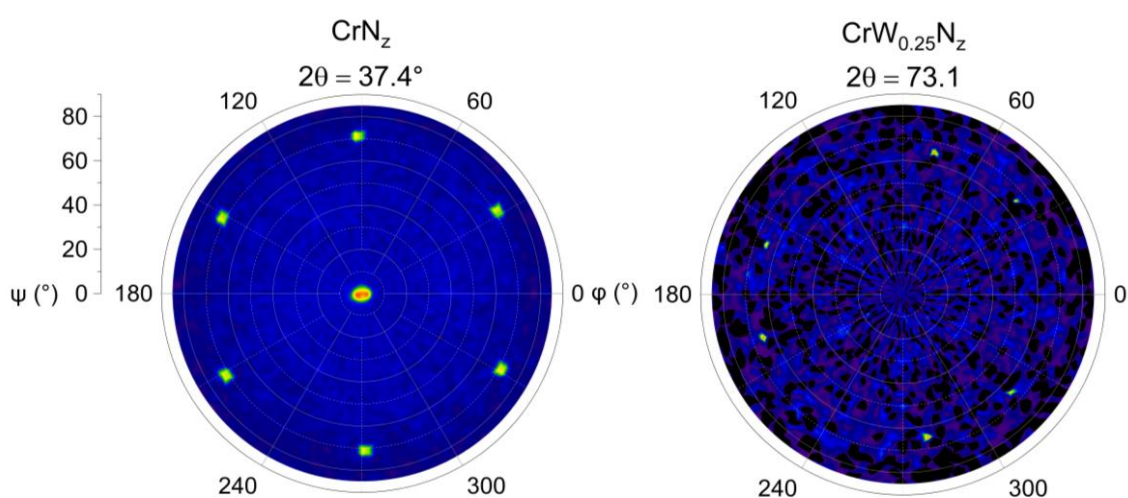

Fig. 4. Pole figure scans from CrN$_z$ and CrW$_{0.25}$N$_z$ film grown on *c*-plane sapphire for (a) {111} reflections at 2θ = 37.4° and (b) {03$\bar{3}$1} reflections at 2θ = 73.1°, respectively.

Analyses of XRR data (See supplementary, Fig. S1) show that all the CrW$_x$N$_z$ films were ~ 60 nm thick with a roughness of ≤ 3 nm, irrespective of the W content. The film density increased from ~ 6.7 to 11.1 g cm$^{-3}$ as the W content was increased from x = 0.03 to x = 0.25. The baseline density for CrN films was ~ 5.8 g cm$^{-3}$ which is ~ 98% of the theoretical density of CrN.

*C. Morphological studies*

W alloying had a significant effect on the CrW$_x$N$_z$ film morphology. Pristine CrN film surface exhibit large triangular grains (Fig. 5(a)) indicative of 111-oriented rocksalt films [43]. In contrast, films with x = 0.03 show highly elongated grains with an average width of ~ 70 nm, however, the average lengths of these grains are difficult to calculate due to continuous and curved domains (Fig. 5(b)). Films with higher W contents show a different morphology indicative of fine *h*-Cr$_2$N grains. However, for x = 0.07, we observe a dense morphology (Fig. 5(c)). Films with x = 0.19 and 0.25 (Fig. 5(d-e)) both exhibit similar features having average grain size ~ 40 nm, but with the presence of voids. Increasing the W content leads to grain refinement, *h*-Cr$_2$N formation and W precipitation. The grain size distribution for the corresponding films is plotted in Fig 5(f-h). Figs. S3 show elemental mapping of films with x = 0.25, showing that the W precipitates are homogeneously distributed throughout the film.

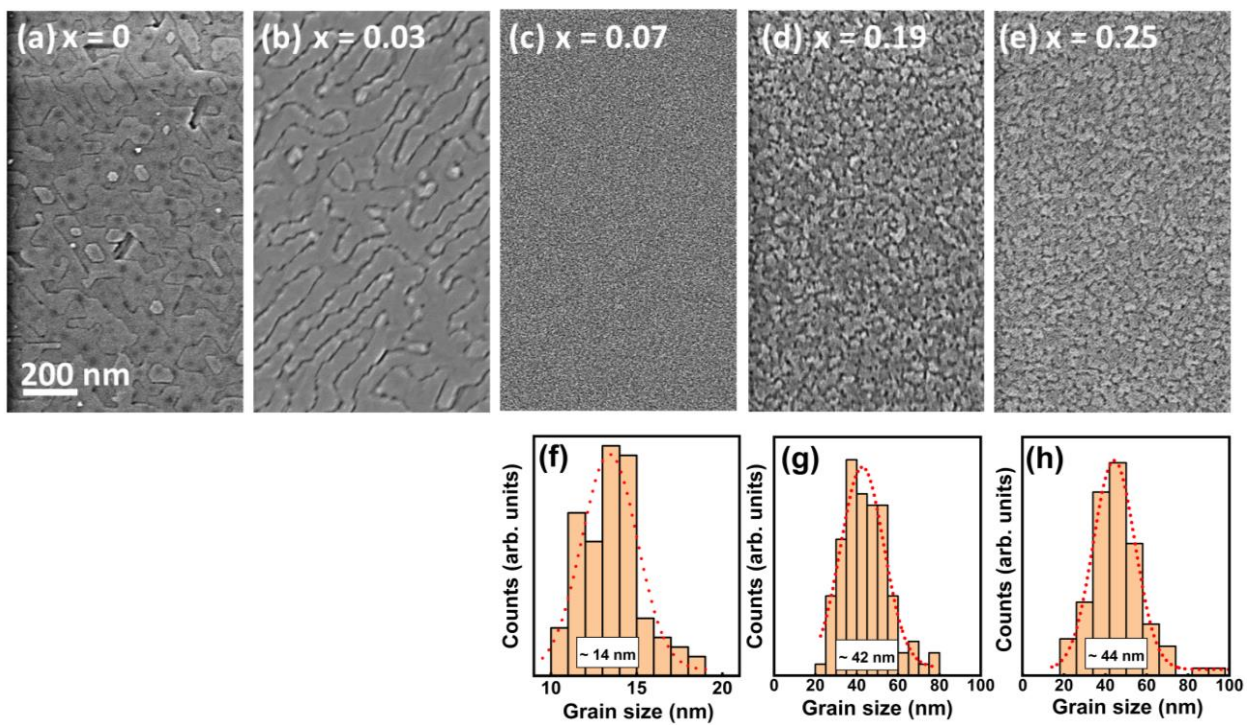

Fig. 5 (a-e) SEM micrographs from $CrW_xN_z$ films with $0.00 \leq x \leq 0.25$, and (f–h) corresponding grain size distribution plot with Gaussian fit.

*C. Thermoelectric transport properties*

The room-temperature values of $\sigma$ and $\alpha$ measured from the films are plotted in Fig 6 as a function W alloying content. Pristine CrN shows $\sigma \sim 197$ Scm$^{-1}$ which agrees with prior reports for CrN films deposited under similar conditions [5]. Adding W increases $\sigma$ to $\sim 700$ Scm$^{-1}$ for $x = 0.03$, which is in accordance with the W-induced changes in the electronic band structure and increase in the density of states at the Fermi level indicated by our *ab initio* calculations. For $x > 0.03$, however, our theoretical calculations do not directly apply because of *h*-$Cr_2N$ phase formation that was not considered in our computations. Films with higher W contents in the $0.07 \leq x \leq 0.19$ range exhibit no considerable changes in $\sigma$ suggesting that the increase in $\sigma$ expected from increased metallic *h*-$Cr_2N$ fraction is countered by the decreased rocksalt CrN fraction. At $x = 0.25$, $\sigma$ increases to $\sim 1800$ Scm$^{-1}$ due to W precipitation [44].

The thermoelectric properties and the Hall measurement analyses are shown in Fig. 5. All the films show negative $\alpha$, indicating n-type behavior, i.e., electrons are the majority charge carriers. Pristine CrN shows high $\alpha \sim -155$ µV/K which reduces greatly to $\alpha \sim -25$ µV/K for $x = 0.03$. Experimental observation of lower $\alpha$ for $x = 0.03$ as contrary to the theoretical calculations infers one needs to explore much lower W content CrN films to avoid formation of secondary phases. With even higher W content ($x > 0.03$), $\alpha$ reduces further and saturates around $\alpha \sim 5$ µV/K due to the increased metallic *h*-$Cr_2N$ phase.

The room-temperature power factor $\alpha^2\sigma \sim 4.7 \times 10^{-4}$ Wm$^{-1}$K$^{-2}$ for CrN is in close agreement as reported by Kerdsongpanya *et al.*, ($\sim 5 \times 10^{-4}$ Wm$^{-1}$K$^{-2}$) [21], but lower than those of Gharavi *et al.*, ($\sim 9$-$11 \times 10^{-3}$ Wm$^{-1}$K$^{-2}$) [45]. However, due to the reduced $\alpha$, all the films with W shows lowered $\alpha^2\sigma$ values.

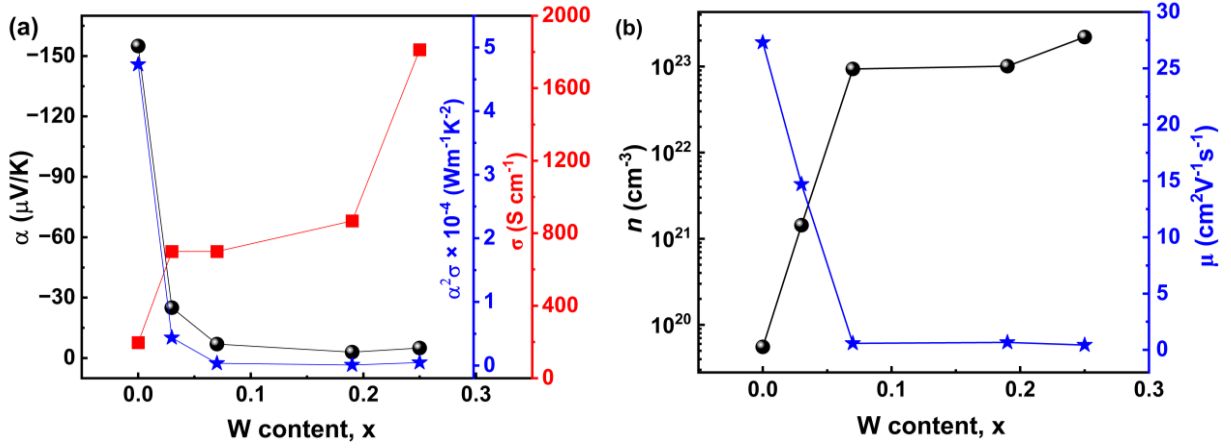

Fig. 6 (a) Seebeck coefficient, electrical conductivity, and corresponding power factor, and (b) carrier density $n$ and mobility $\mu$ extracted from Hall measurements in $CrW_xN_z$ films.

Analyses of Hall measurements of the $CrW_xN_z$ films (Fig. 5 (b)) indicate W additions result in a up to four-orders-of-magnitude increase in charge carrier density n, accompanied by a two-orders of magnitude decrease in carrier mobility µ. The large increase in carrier density are due to metallic $h$-$Cr_2N$ [46] and W precipitation. The measured $n \sim 5.5 \times 10^{19}$ cm$^{-3}$ for CrN agrees well with prior reports [21]. The low µ is typical of metals and is attributable to grain-boundary and W impurity scattering mechanisms. Since $\alpha$ and $\sigma$ are strongly and adversely coupled through $n$, the monotonic rise in $n$ increases $\sigma$ but decreases $\alpha$, leading to a net decrease in $\alpha^2\sigma$. Thus, it is crucial to restrict W additions to below its solubility limit in CrN to preclude the formation of $h$-$Cr_2N$ and W precipitation to realize high $\alpha^2\sigma$.

## IV. Conclusions

*Ab initio* calculations indicate that dilute W additions to CrN are attractive to increase the thermoelectric power factor due to the emergence of flat bands near the Fermi level pushed into the conduction band. Epitaxial growth of dc-magnetron sputtered W alloyed CrN films showed metallic $h$-$Cr_2N$ formation even for the smallest W contents explored in this study. Despite this, the power factor is considerable at x = 0.08 due to high $\sigma \sim$ 700 Scm$^{-1}$ and sizeable $\alpha$ = -25 μV/K. Increasing W content severely diminishes α despite the monotonic increase in σ due to increasing $h$-$Cr_2N$ fraction and W precipitation. These findings indicate that limiting W alloying to below its solubility limit in CrN is crucial for realizing a high power factor.

**Conflict of interest**

The authors declare no conflict of interest.

**Data availability**

Data is available from the corresponding author on reasonable request.

**Acknowledgements**

The authors acknowledge funding from the Swedish Government Strategic Research Area in Materials Science on Functional Materials at Linköping University (Faculty Grant SFO-Mat-LiU No. 2009 00971), the Knut and Alice Wallenberg foundation through the Wallenberg Academy Fellows program (KAW-2020.0196), the Swedish Research Council (VR) under Project No. 2021-03826, and the Swedish Energy Agency under project number 52740-1. NKS and PE acknowledge support from the Foundation Olle Engkvist Byggmästare, grant no. 218-0076. NKS acknowledges Dr Rui Shu for his help during the initial experimental phase. GR gratefully acknowledges funding from the US National Science Foundation grant CMMI 2135725 through the BRITE program. DG acknowledges support from the Swedish Research Council (VR) International Postdoc Grant 2023-00208. BA acknowledges financial support from the Swedish Research Council (VR) through Grant No. 2019-05403 and 2023-05194. The computations were enabled by resources provided by the National Academic Infrastructure for Supercomputing in Sweden (NAISS), at the National Supercomputer Center (NSC), partially funded by the Swedish Research Council through grant agreements no. 2022-06725.

Supplemental material

# Effects of W alloying on the electronic structure, phase stability and thermoelectric power factor in epitaxial CrN thin films

Niraj Kumar Singh,[1,2*] Victor Hjort,[1] Sanath Kumar Honnali,[1] Davide Gambino,[3] Arnaud le Febvrier,[1] Ganpati Ramanath[1,3], Björn Alling[3] and Per Eklund[1,2]

[1]*Thin Film Physics Division, Department of Physics, Chemistry, and Biology (IFM), Linköping University, SE-581 83, Linköping, Sweden*

[2] *Inorganic Chemistry, Department of Chemistry - Ångström Laboratory, Uppsala University, Box 538, SE-751 21 Uppsala, Sweden*

[3]*Theoretical Physics Division, Department of Physics, Chemistry, and Biology (IFM), Linköping University, Linköping-58183, Sweden*

[4]*Rensselaer Polytechnic Institute, Materials Science & Engineering Department, Troy, NY 12180, USA*

*E-mail: niraj-kumar.singh@kemi.uu.se

This supplemental material presents of details of CrN film growth as a function of Cr target power $P_{Cr}$, analyses of X-ray reflectograms from $CrW_xN_z$ films to determine fim thickness, roughness and density, and EDX maps of films with the highest W content.

Prior to depositing films with different W contents, CrN films were deposited at different Cr target power $P_{Cr}$ to preclude the formation of $Cr_2N$ formation and Cr-deficient films. X-ray diffractograms from films deposited with 42 W ≤ $P_{Cr}$ ≤ 46 W (Fig. S1a) show 111 and 222 Bragg peaks from rocksalt CrN indicating a lattice parameter of $a_0$ ~ 4.17 Å [9]. The 0006 peak is from the sapphire substrate. At $P_{Cr}$ = 50 W, we observe large 0002 peaks from *h*-$Cr_2N$ corresponding to c ~ 4.48 Å, and a weak CrN 111 peak. Among these, CrN film grown at $P_{Cr}$ = 44 W showed the highest thermoelectric power factor, $\alpha^2\sigma$ (Fig. S1b).

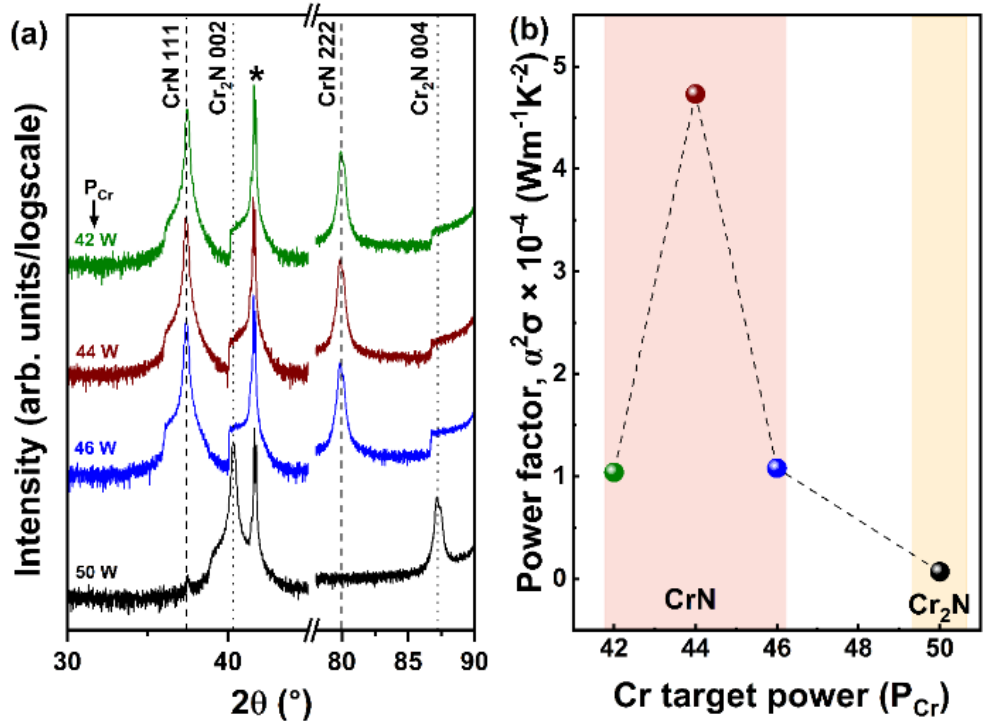

Fig. S1. (a) X-ray diffractograms, and (b) thermoelectric power factor, $\alpha^2\sigma$ of CrN films as a function of Cr target power $P_{Cr}$.

The XRR data was fitted to extract the film thickness, roughness and density (see Fig. S2).

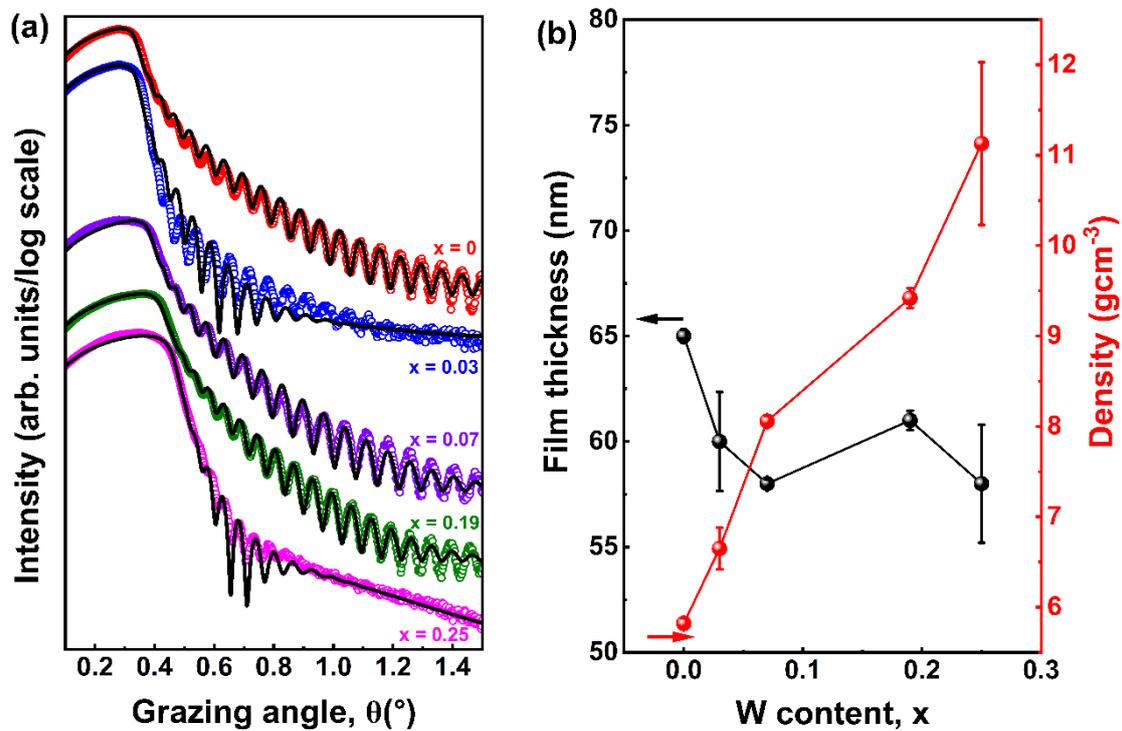

Fig. S2. (a) Measured XRR data and fits for $CrW_xN_z$ ($0 \leq x \leq 0.25$) films and (b) the extracted values of film thickness and density.

$CrW_xN_z$ films with the highest W content of x=0.25 showed homogenous contrast from elemental W indicating homogeneous distribution of W and Cr (Fig. S3).

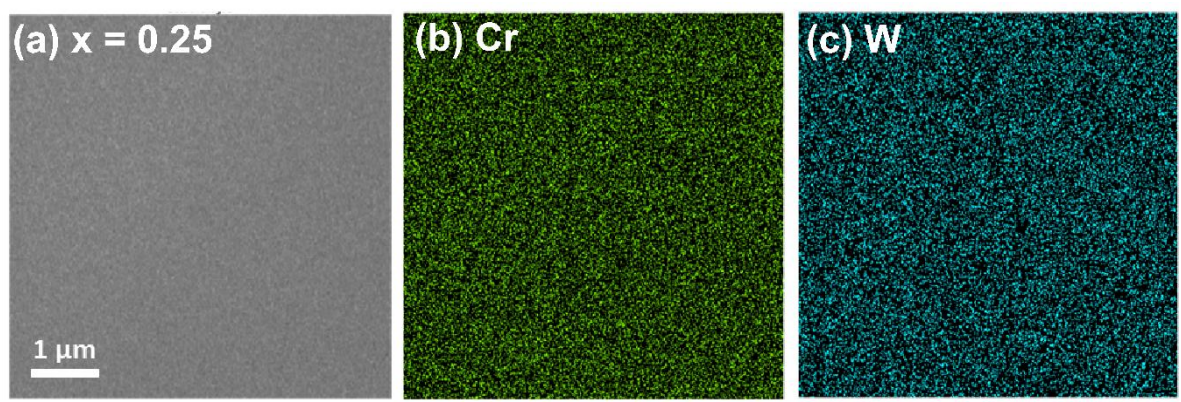

Fig. S3. (a-c) Elemental EDX maps of Cr and W from $CrW_{0.25}N_z$ films.